# Review of the 10th Non-LTE Code Comparison Workshop

S.B. Hansen[1], H.-K. Chung[2], C.J. Fontes[3], Yu. Ralchenko[4], H.A. Scott[5], E. Stambulchik[6]

[1]Sandia National Laboratories, Albuquerque, NM 87185, USA
[2] Gwangju Institute of Science and Technology, Gwangju, 61005, Republic of Korea
[3]National Institute of Standards and Technology, Gaithersburg, MD 20899-8422, USA
[4]Lawrence Livermore National Laboratory, Livermore, CA 94550, USA
[5]Faculty of Physics, Weizmann Institute of Science, Rehovot 7610001, Israel

**Abstract**

We report on the results of the 10th Non-LTE code comparison workshop, which was held at the University of San Diego campus November 28 through December 1, 2017. Non-equilibrium collisional-radiative models predict the electronic state populations and attendant emission and absorption characteristics of hot, dense matter and are used to help design and diagnose high-energy-density experiments. At this workshop, fifteen codes from eleven institutions contributed results for steady-state and time-dependent neon, aluminum, silicon, and chlorine cases relevant to a variety of high-density experimental and radiation-driven astrophysical systems. This report focuses on differences in the predictions from codes with different internal structure, completeness, density effects, and rate fidelity and the impact of those differences on hot, dense plasma diagnostics.

**Introduction**

The assumption of Local Thermodynamic Equilibrium (LTE) provides a relatively simple picture of the statistical properties of material at extreme conditions. In LTE, electronic state occupations within a single charge state are directly determined by statistical weights and temperature-dependent Boltzmann factors; and the populations of adjacent ionization stages are determined by temperature- and density-dependent Saha-Boltzmann factors. While this LTE description can be adequate for long-lived, high-energy-density plasmas (HEDP) whose internal transitions are dominated by atomic processes in strict equilibrium, a multiplicity of departures from LTE are possible, and indeed most laboratory and many astrophysical plasmas are non-LTE. In such cases the electronic state occupations must be determined using a set of coupled collisional-radiative (CR) rate equations that may or may not be explicitly time-dependent [1]. The resultant model complexity has stimulated a series of code comparison workshops that aim to validate and, where benchmark data exist, verify the collisional-radiative models used in the astrophysical and HEDP communities. This report describes the results of the tenth such workshop.

Previous workshops have explored both steady-state and time-dependent calculations of fusion-relevant materials from carbon to gold [2-8]. Specific cases have explored external radiation fields, optical depth, non-Maxwellian electrons, models constrained by common structure, specified exclusion of particular processes, and cases tailored to various applications such as hohlraum emission and plasma diagnostics. These studies and the increasingly sophisticated database and analysis tools developed for the workshops have helped the collisional-radiative modeling community systematically determine the basic – and application-dependent – requirements for reliable non-LTE modeling [1]. Among these requirements are a) state-space completeness, especially in autoionizing and multiply excited levels, for accurate charge state and radiative loss calculations and b) detailed atomic structure coupled with accurate rate data for detailed spectroscopic diagnostics. Generally reliable models must balance these requirements with computational tractability, which becomes increasingly difficult with both completeness and detail of the models. They must also include adequate treatments of the density effects that impact model predictions in HEDP regimes relevant to warm dense matter (WDM) and inertial confinement fusion (ICF). These density effects include continuum lowering, level destruction, collisional population redistribution, and line broadening, none of which are based on completely settled theory and all of which can be implemented in diverse ways in CR models. This fundamental uncertainty in high-density regimes – and its impact on plasma diagnostics relevant to recent experiments – provided motivation for many of the cases explored in the 10$^{th}$ non-LTE workshop.

**Workshop organization**

The Non-LTE code comparison workshops are held every two years, alternating between U.S. and European locations. The tenth workshop was held on the campus of the University of California, San Diego (UCSD), from November 28 through December 1, 2017. Tentative cases were communicated to potential participants in June of 2017 and the call for submissions was finalized several weeks later with supplementary information and a detailed description of the submission format. Contributors were asked to submit results three weeks before the workshop. The results were made available in an online database one week before the workshop to give case coordinators time to look through the collected results and contributors time to check and, if desired, amend their submissions. As summarized in Table I, 36 participants from 6 countries and 11 institutions contributed results from more than 26 variations of 17 independent codes, with variations arising from the choices of atomic data and/or models for density effects. Only 21 scientists attended in person, in part due to visa issues.

The first days of the workshop were devoted to a combination of code talks and invited talks. The code talks provided opportunities for an author or representative of each code to describe its structure, status, recent changes, and applications. Invited speakers on topics relevant to non-LTE experiments and observations included Farhat Beg of UCSD on current experimental efforts at the Center for Energy Research, G. Loisel of SNL on benchmark spectroscopic measurements from plasmas photoionized by x-rays from Sandia's Z-machine, E. Marley of LLNL on the development of buried-layer platforms to study spectra from non-LTE plasmas at Rochester's Omega laser, and M. MacDonald of LLNL on spatially-resolved x-ray fluorescence measurements from Omega and Livermore's National Ignition Facility (NIF). The last days of the workshop were devoted to the detailed case discussions that comprise the heart of these workshops.

As in past workshops of this series, only contributors and invited speakers attended the workshop. While anyone is welcome to contribute calculations, only contributors have access to the full database and only anonymized data are made publicly available. These restrictions encourage contributors to submit complete data sets without concern that their codes will be called out as outliers, which stimulates open discussion and reduces the risk of artificial consensus.

Table I. Codes, Contributors, and submissions.

| Code | Contributors | Institute | Country | Ne | Al | Si | Cl | Ne TD | Al TD |
|---|---|---|---|---|---|---|---|---|---|
| ACRE [9] | R. Abrantes | UCLA | USA | | | | X | | |
| ABAKO [10] | R. Florido, J.M. Martin-Gonzalez, M.A. Gigosos | ULPGC | Spain | X | | | | | |
| ATLANTIS [11] | M.A. Mendoza, J.G. Rubiano, R. Florido, J.M. Gil, R. Rodriguez, P. Martel, A. Benita, E. Minguez | ULPGC | Spain | X | X | X | X | | |
| ATOMIC [12] | C. Fontes, J. Colgan, M. Zammit | LANL | USA | X | X | X | X | X | |
| CORD [13] | M. Poirier | CEA | France | X | | | X | | |
| CRAC | E. Stambulchik | Weizmann Inst Sci | Israel | X | X | X | X | X | X |
| CRETIN [14] | H. Scott, P. Grabowski, H. Le | LLNL | USA | X | X | X | X | X | X |
| DEDALE [15] | F. Gilleron, R. Piron, M. Comet, J.-C. Pain | CEA | France | X | X | | X | | |
| DLAYZ [16] | G. Cheng, Z. Jiaolong, Y. Jianmin | NUDT | China | X | X | X | X | X | X |
| DRACHMA2 [17] | N. Ouart | NRL | USA | | | | X | | |
| NOMAD [18] | Dipti, Y. Ralchenko | NIST | USA | X | | X | X | X | |
| OPAZ [19] | C. Blancard | CEA | France | X | X | X | X | | |
| PrismSPECT [20] | I. Golovkin | Prism Comp. | USA | X | X | X | X | | |
| SCRAM [21,22] | S. Hansen, B. Kraus | SNL | USA | X | X | X | X | | X |
| SCSF [23] | S. Hansen | SNL | USA | X | X | X | X | X | X |
| SEMILLAC [24,25] | Y. Frank | LLNL | USA | X | X | X | X | | |
| THERMOS [26] | I. Vichev, D. Kim, A. Solomyannaya | KIAM | Russia | X | X | X | X | | |

**Cases**

The cases for this workshop, summarized in Table II, were selected for their relevance to current experiments. The steady-state Ne, Al, and Cl cases aimed to explore the collisional-radiative kinetics and K-shell emission signatures of elements that are used as diagnostic tracers in HEDP experiments. Spanning a wide range of densities, these cases allowed us to investigate density-dependent effects such as intercombination line intensities, line broadening, degeneracy-driven changes to rates, and continuum lowering. Contributors were encouraged to submit model variations to help isolate these effects. The Al and Ne cases also served as a baseline for the time-dependent cases described below. Contributors were encouraged to submit best-fit Cl spectra for comparison with recent high-resolution measurements from the OHREX spectrometer fielded at the Orion facility [27].

The steady-state Si cases were designed to assess model variations for astrophysically relevant plasmas whose ionization is dominated by radiative rather than collisional excitation and ionization. Results from these cases were compared with benchmark data collected from a photoionized plasma at Sandia's Z facility [28], where the observed charge state distribution of the experimental photoionized plasma has not yet been reproduced by collisional-radiative models at the measured plasma conditions. The external radiation field was characterized by either a single Planckian or a sum of three Planckians selected to fit the measured driving radiation field. Multiple plasma lengths were specified to explore model predictions relevant to Resonant Auger Destruction [29-31], a proposed mechanism by which K-shell satellite lines from L-shell ions are suppressed as they are transported due to dominant Auger decay rates that inhibit direct re-emission following resonant absorption.

Finally, time-dependent Al and Ne cases were specified to revisit, with more realistic initial conditions and fixed temporal dependence, the time-dependent argon case explored in NLTE-7 relevant to X-ray free-electron laser (XFEL) experiments, which showed strong dependence on model completeness. The Al case was relevant to the experiment described in [32,33] and follow-on publications, where time-integrated fluorescence emission as modeled by SCFLY [34] called into question standard Stewart-Pyatt models of continuum lowering, and the low-density Ne case was relevant to the multi-step ionization measurements described in [35].

In the following sections we present a summary of the results from the workshop for these cases as well as a discussion of the implications for diagnostic interpretation and the relevance of the results to the experiments that inspired each case. For each case, we present global results such as average charge (Z*) and total radiative power loss rates (RPL) along with diagnostically relevant information such as line ratios and detailed spectra. In all figures, the line styles reveal two aspects of the models: the weight of each line (thick or thin) corresponds to the statistical completeness of each model (as represented by its total statistical weight), while the darkness of the line corresponds to the finest level of electronic-state detail. So, for example, a statistically complete model based on superconfigurations will be represented by a thick, light line, while a fine-structure model with a relatively small total statistical weight will be represented by a thin dark line. In general, the more statistically complete models are more reliable predictors of global values like Z* and RPL, while more detailed models are better suited to spectroscopic diagnostics.

Table II. Cases specified for the NLTE-10 workshop. Along with the call for resubmission, additional temperatures and an additional radiation field were requested for Si, indicated with brackets.

| Element | Case ID | No. of cases | $T_e$ (eV) | $N_e$ (cm$^{-3}$) | $T_{rad}$ (eV), dilution factor | Plasma radius (cm) | Spectral ranges |
|---|---|---|---|---|---|---|---|
| Ne | Ne | 12 | 50, 100, 200, 500 | $10^{19}, 10^{20}, 10^{21}$ | | | 800-1400 eV, $\delta\varepsilon$=0.3 eV |
| Ne | Ne-TD | 3 | $T_e(t)$ for 3 cases in supplemental file* | $N_e(t) = Z^*(t) \times N_i$ with $N_i=10^{18}$ | $E_{rad}$: 800, 1050, 2000 | | 800-1400 eV, $\delta\varepsilon$=0.3 eV |
| Al | Al | 8 | 10, 30, 100, 300 | $2\times10^{23}, 5\times10^{23}$ | | | |
| Al | Al-TD | 2 | $T_e(t)$ for 2 cases in supplemental file* | $N_e(t) = Z^*(t) \times N_i$ with $N_i=6\times10^{22}$ | $E_{rad}$: 1580, 1650 | | 1400-2400 eV, $\delta\varepsilon$=0.5 eV |
| Si | Si | 12 [24] | 30 [60, 100] | $10^{19}, 3\times10^{19}$ | $T_{rad}$: 63 [diluted], multi-Planckian | 0.1, 0.3, 1.2 | 1700-2500 eV, $\delta\varepsilon$=0.25 eV |
| Cl | Cl | 9 | 400, 500, 600 | $10^{21}, 10^{22}, 10^{23}$ | | | 2600-3800 eV, $\delta\varepsilon$=0.15 eV |

**Results**

Steady-state Ne

Neon is the fourth-most abundant element in the sun, contributing about 10% to the total solar opacity. Its K-shell lines are useful diagnostics for moderate-temperature HED plasmas [36,37], and as a noble gas it finds many applications as an absorber and/or radiator in gas targets. The steady-state neon cases were well represented in the workshop, with 26 variations of 15 independent models submitted. The models had wide variations in statistical completeness, with total model statistical weights ranging from $10^3$ to $10^8$. About 30% of the submitted models were highly averaged (hydrogenic superconfigurations) while the rest included configuration splitting and/or fine-structure states.

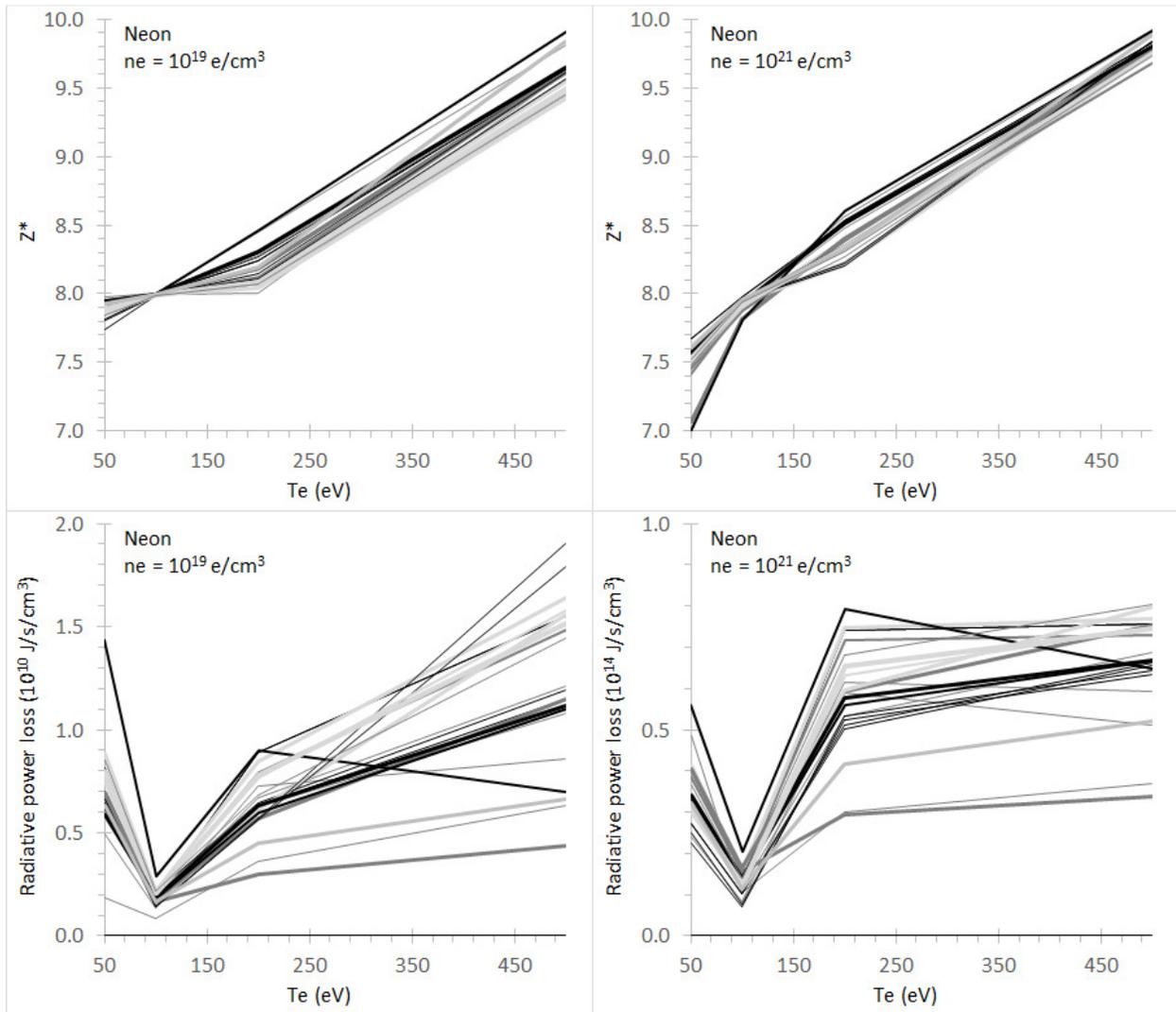

Figure 1. Average ionization $Z^*$ and radiative power loss rates of neon.

Figure 1 shows the temperature dependence of the average ion charge $Z^*$ and radiative power loss rates at the highest and lowest case densities. As in previous workshops, the codes tend to agree best near closed-shell configurations (for neon, the closed-shell He-like ion has $Z^* = 8$). The more statistically complete but highly averaged models (thick, light-gray lines) tend to have weaker temperature dependence in $Z^*$ around the closed shell and larger radiative loss rates than the more detailed models. Notably, there are clusters of results that sample most (but not all!) of the models that are both relatively detailed and relatively complete (thicker black lines) in all four plots of Fig. 1. These models used several different sources for their atomic data, so the trends should be robust.

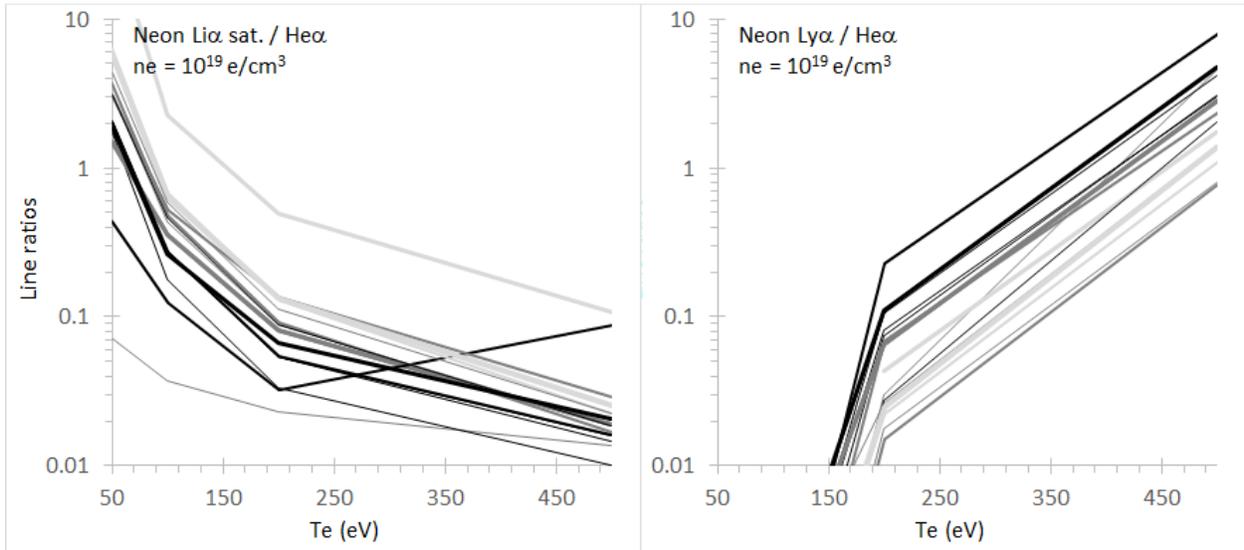

Figure 2. Temperature-sensitive line ratios in neon.

Figure 2 provides a glimpse into the practical implications of how differences in the modeled $Z^*$ might affect temperature diagnostics based on line ratios from adjacent charge states. Ratios of the Li-like satellite to He$\alpha$ resonance line intensities are illustrated on the left of Fig. 2. While the differences among all models are factors of 10 -100, differences among a cluster of relatively complete and relatively detailed models (thicker and darker lines) are only factors of ~2. Similar trends are evident in the Ly$\alpha$ to He$\alpha$ line intensity ratios given on the right.

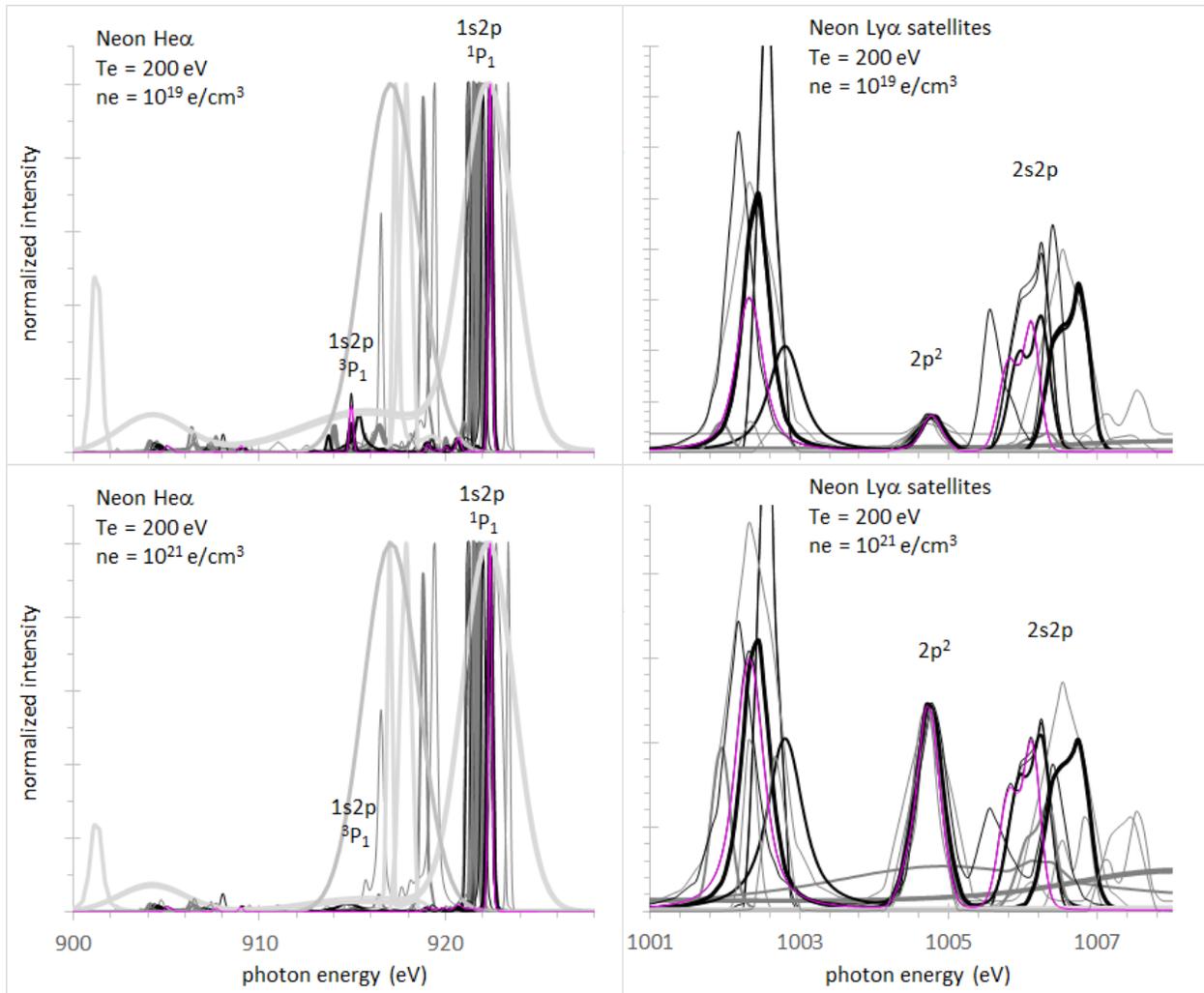

Figure 3. Left: The density dependence of the He-like neon intercombination line and Li-like satellites from all models, intensity-normalized to the resonance line. Right: the density dependence of He-like satellites to the H-like neon Lyα lines from relatively detailed models, which have been shifted in energy to align the $2p^2$ satellite feature.

Intensity-normalized spectra near Heα and Lyα lines are given in Fig. 3, illustrating both model variations in emission line energies and structure and variations in two density-sensitive line ratios that can be used to diagnose plasma densities [36]. On all plots, a single fine-structure model has been called out in magenta to help guide the eye. In the Heα spectra shown at right, the relatively detailed models (black lines) are more tightly clustered together than the highly averaged models (gray lines). Among the most detailed models, the emission line energies disagree by about 2 eV and among all models the variation reached 10 eV. All of the detailed (and a few of the modestly averaged) models predict a decrease in the intensity of the Heα intercombination line ($^3P_1$) relative to the Heα resonance ($^1P_1$) line with increasing density. However, the magnitude of that decrease differs among those models by about a factor of two due to different treatments of the collisional rates that couple excited states in the He-like ion. It is these rates that destroy the metastable-state populations that give rise to the intercombination line. At left, the $2p^2$ feature of the satellite emission to Lyα have been aligned in both wavelength and intensity for clarity (energy shifts up to 3 eV). Even with this alignment there remains a few-eV variation in the emission energies of the other satellite features. All but three of the relatively detailed models show a marked increase in the intensity of the $2p^2$ feature relative to the 2s2p doublet with density, as increasing collisions redistribute population among autoionizing states in the He-like ions. Among the most detailed models (black lines) the quantitative agreement in the $2p^2$/2s2p ratios is within 50% over this density. Between $10^{20}$ and $10e^{21}$ e/cm$^3$, these ratios are in good agreement with those published in [37], supporting its use as a density diagnostic.

**Time-dependent Ne**

The time-dependent Ne case was designed to explore how models handle the dynamic ionization processes experimentally observed in [35]. There, a relatively dilute neon gas was irradiated by X-ray free-electron laser (XFEL) beams of various photon energies. At 800 eV, the photons have enough energy to photoionize only the valence L-shell electrons of neon, and a smooth distribution of charge states up to about Ne$^{6+}$ was observed. At 1050 eV, the beam can photoionize inner-K-shell electrons from ions up to Ne$^{6+}$, producing highly-excited states in Ne$^{(X+1)+}$ ions that can undergo Auger decay to produce Ne$^{(X+2)+}$. Ion distributions measured from these experiments had distinct deficits of odd-numbered charge states including Ne$^{1+}$, Ne$^{3+}$, and Ne$^{5+}$, with few ions above Ne$^{8+}$. Finally, at a beam energy of 2000 eV, the beam can ionize the K-shell electrons of all neon ions; the measured distribution showed slightly less pronounced deficits in odd-numbered charge states than with the beam at 1050 eV and significant populations in ions up to Ne$^{10+}$.

Many fewer codes provided results for the time-dependent cases than for the steady-state cases: there were a total of nine submissions including variations of six unique codes, most of which were highly averaged. The predicted $Z^*(t)$, charge state distributions at two times, and total emission spectra for two cases are given in Fig. 4. The 800 and 2000 eV XFEL beams were modeled as flat-top pulses with, respectively, fluxes of 2.35 and 3.48 x $10^{17}$ W/cm$^2$ and durations of 340 and 230 fs (ending about when $Z^*$ stops increasing in the top panels of Fig. 4). For both cases, the beams had a bandwidth of 4 eV, the ion density remained constant at $10^{18}$ e/cm$^3$ and the electron temperatures, which began at 1 eV and ended at 404 eV for the 800 eV beam and at 621 eV for the 2000 eV beam, were specified by SCFLY [34].

For both XFEL energies, the code predictions for the initial steady-state $Z^*$ at 1 eV spanned several orders of magnitude. Some codes predicted an immediate onset of a near-linear ionization increase while others reached a linear ionization stage only after ~$10^{-16}$ s. A gap in the specified time steps (illustrated by the dots on the inset $Z^*$ curves) led to some sharp changes in the predictions of several codes, and some contributions used individual interpolations that may have affected their predictions. The final $Z^*$ values spanned several charge states, which are not the collisional-radiative equilibrium values at the final temperatures. Reaching true steady-state equilibrium takes much longer than the duration modeled here since the collisional ionization rates from the high charge states after the beam is off (~ $10^{10}$ s$^{-1}$) are much slower than the photoionization rates during the XFEL pulse (~$10^{13}$ s$^{-1}$).

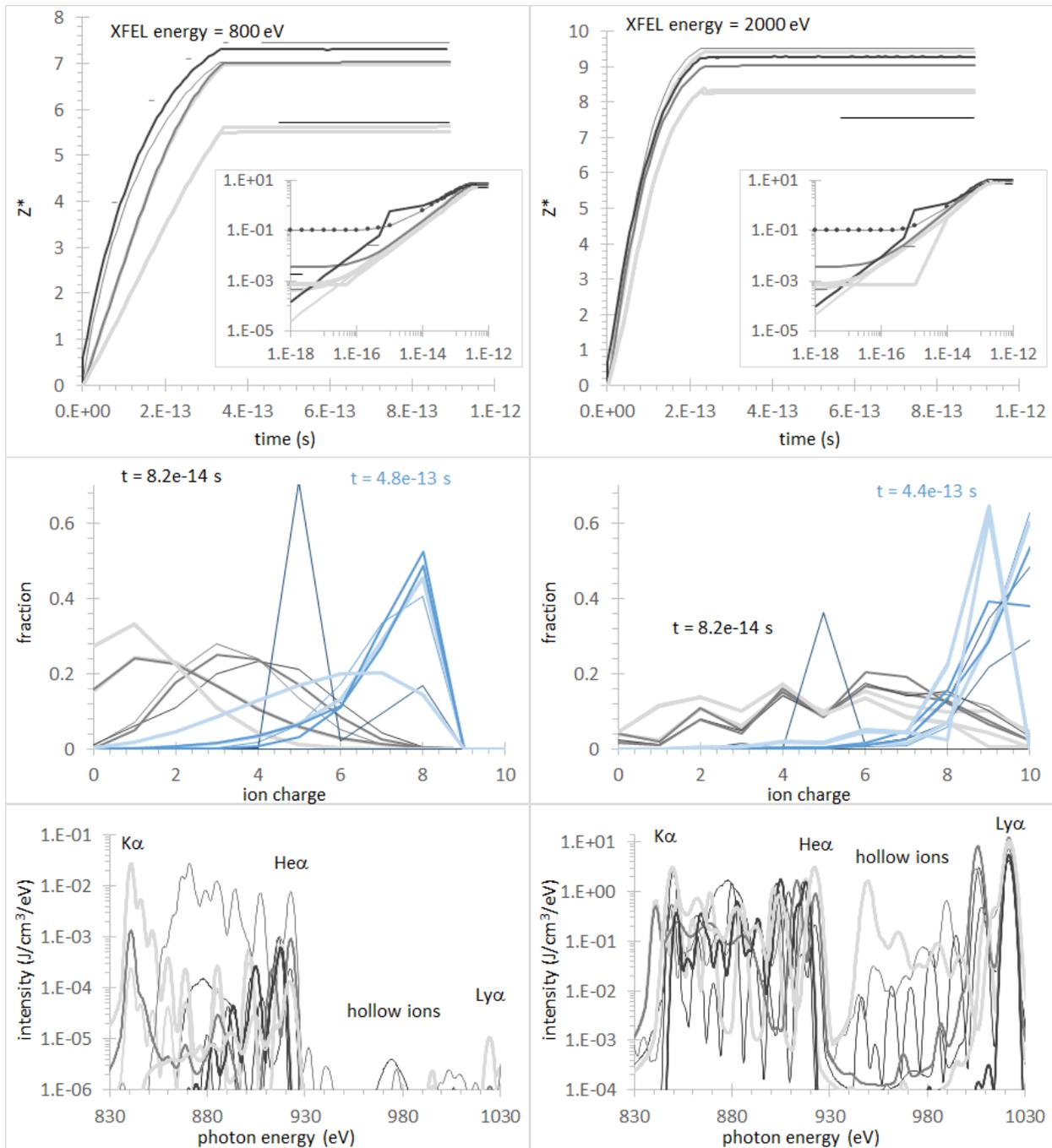

Fig 4. (color online) Top: average ion charge as a function of time for low-density neon irradiated with 800 eV and 2000 eV x-ray free-electron laser beams. Middle: charge state distributions at early (black) and late (blue) times. Bottom: time-integrated emission spectra including 2 eV broadening for clarity.

The charge state distributions shown in the middle panels of Fig. 4 are consistent with the expected ionization mechanisms of the two XFEL cases. With a beam energy of 800 eV, the dominant process is sequential L-shell photoionization and the charge state distributions are smooth, both during (black and gray lines) and after (blue shaded lines) the XFEL pulse with the exception of an outlier code that used different time steps. By contrast, the dominant process for the 2000 eV XFEL is K-shell photoionization followed by Auger decay, which gives rise in most of the codes to the distinctive population deficits in odd numbered charge states that were observed experimentally [35]. And as in the experiment, higher charge states are reached with the higher XFEL energy. Aside from several outliers (a relatively detailed code with idiosyncratic time steps and a highly averaged code with idiosyncratic photoionization rates for the first two charge states), the agreement among codes in predicted charge state evolution is quite good for both XFEL energies.

Despite the agreement among most codes in the predicted charge state evolution, there are significant differences in the predicted photoionization emission, as illustrated in the bottom of Fig. 4. For the 2000 eV XFEL beam, most codes predict a relatively uniform distribution of emission from all charge states, but there is a wide variation in predictions for the intensity of emission from hollow ions (ions with two K-shell vacancies) that is roughly correlated with model completeness. For the 800 eV case, the predicted emission intensities vary by orders of magnitude and the spectra can be dominated by near-neutral lines, He-like lines, or some charge state in between. These trends in the emission "fingerprint" are not clearly correlated with differences in the charge state evolution of the models. And it is at first blush curious that the 800 eV case should produce any fluorescence emission at all, since resonant K-shell excitation and ionization energies (848 and 870 eV, respectively, for neutral neon) lie above the beam energy for *all* neon ions. The density is too low for continuum lowering/ionization potential depression to allow direct photoionization, and while there are ~10 eV differences in ionization potentials and K$\alpha$ energies among models, these are not sufficient to allow direct photoexcitation from a square beam. Yet all codes predict K-shell emission many orders of magnitude larger than would arise from collisional processes at the prescribed temperatures, and only a few orders of magnitude lower than K-shell emission from the 2000 keV beam case. The K-shell holes that give rise to emission above the XFEL beam energy are caused by dielectronic recombination through $(1)^2(n…N)^X + e^- \rightarrow (1)^1(n…N)^{X+2}$ channels, a process whose magnitude varies widely among the codes and is strongly correlated with the predicted emission intensity for each charge state.

**Steady-state Al**

Aluminum is a common component of high-energy-density experiments, with K-shell emission that can help diagnose plasmas with temperatures below about 1 keV. The cases selected for aluminum in this workshop were designed to test density effects such as line broadening, continuum lowering, and degeneracy effects on collisional rates, and helped establish baseline model parameters for the more complex time-dependent cases relevant to the experiments of Vinko et al [33]. Results were submitted for 26 variations of 12 independent codes; the variations primarily concerned different continuum lowering treatments. In the results below, models without continuum lowering effects are called out in red, models using the standard Stewart-Pyatt and/or ion sphere treatments are given in grayscale, and models using Ecker-Kroll (or modifications thereof) are shown in shades of blue. Implementations of continuum lowering in most modern non-LTE models are fairly ad-hoc and have, in general, two components. First is a reduction in ionization potential energies following various theories (Stewart-Pyatt, ion-sphere, or, more recently, Ecker-Kroll). Second is a reduction of statistical weight to gradually move bound states in to the continuum. All of the models that included continuum lowering here include the first component and all but two included the second. The two models that excluded state destruction are called out by dashed lines. Most of the submitted models used data averaged to configurations or superconfigurations: only three included fine-structure levels. However, since high densities truncate the statistical state space that must be included for a relatively complete model, most codes had relatively high completeness, with total statistical weights ranging from $10^3$ to $10^6$.

Figure 5 shows the average ion charge as a function of temperature for the two steady-state case densities. The variations in Z* of 2-3 charge states are significantly larger than were seen for neon due to both the complexity of mid-L-shell ions compared to the K-shell neon ions and the profound density effects that control Z* near the closed-shell neon-like charge state. To orient the reader, the L-shell ionization potentials of the first few charge states of aluminum beyond neon ($Al^{3+}$ - $Al^{7+}$) are roughly 100-300 eV for isolated ions, subshell splitting in the n=3 M-shell is roughly 10 eV, and the configuration interaction effects that inform fine-structure splitting are of order 1 eV. For solid-density material, Stewart-Pyatt and ion-sphere theories predict continuum lowering energies of 50 – 100 eV while Ecker-Kroll predicts larger values of 70 -200 eV. Since both the ionization potential and the ionization potential depression have similar magnitudes, small changes in either can lead to large changes in the model predictions.

The Z* results are roughly clustered according to the model treatment of continuum lowering effects: models without continuum lowering have lower Z* values while models using Ecker-Kroll reach higher charge states. It is notable that all but two of the models that include continuum lowering predict Z*=3 at the lowest temperature and density, indicating complete destruction of the M-shell states in the solid, consistent with a picture of aluminum in its cold metallic state as a simple free-electron mental with Z*=3 ($n_e$ =1.8x$10^{23}$ e/$cm^3$). The two models that fall below Z* = 3 both include fine structure, which allows relativistic subshells to be destroyed (or not, in the case of the dashed-line models, which exclude state destruction) at different densities. At higher temperatures, even the models with similar treatments of continuum lowering diverge significantly. Finally, we note that while a partially degenerate electron energy distribution can significantly reduce collisional rates [1,38,39] when temperatures are near the Fermi energy (12 – 16 eV for the two cases), the impact is not evident here because degenerate rates still follow the detailed balance relations that enforce LTE populations, and in steady-state without an external radiation field the lower-temperature cases here are firmly in LTE.

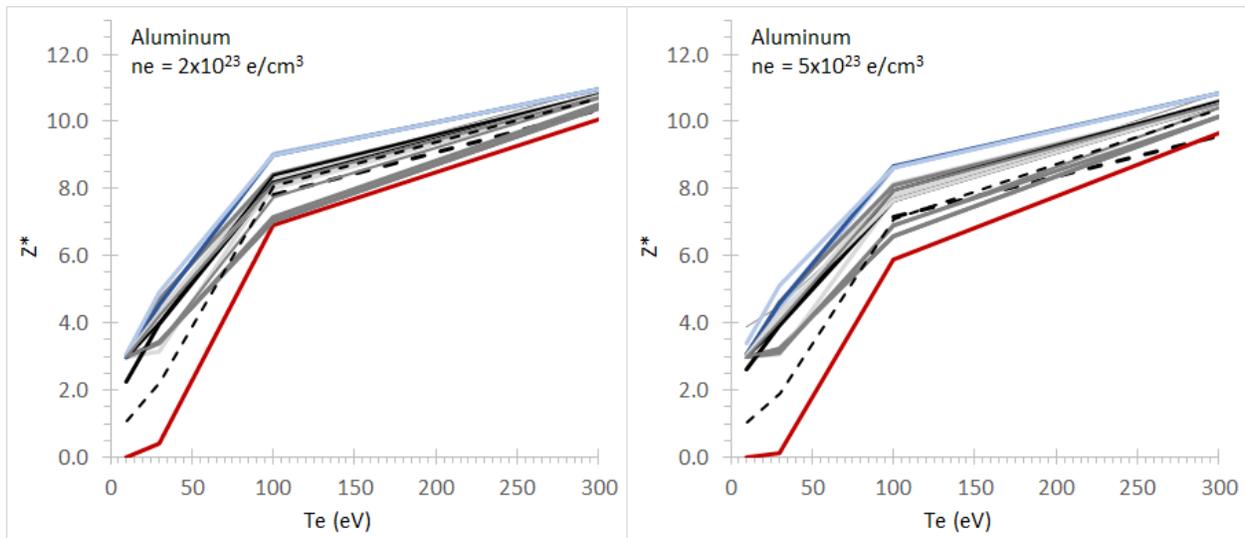

Figure 5. (color online) Temperature dependence of the average ion charge of steady-state aluminum at near-solid electron densities.

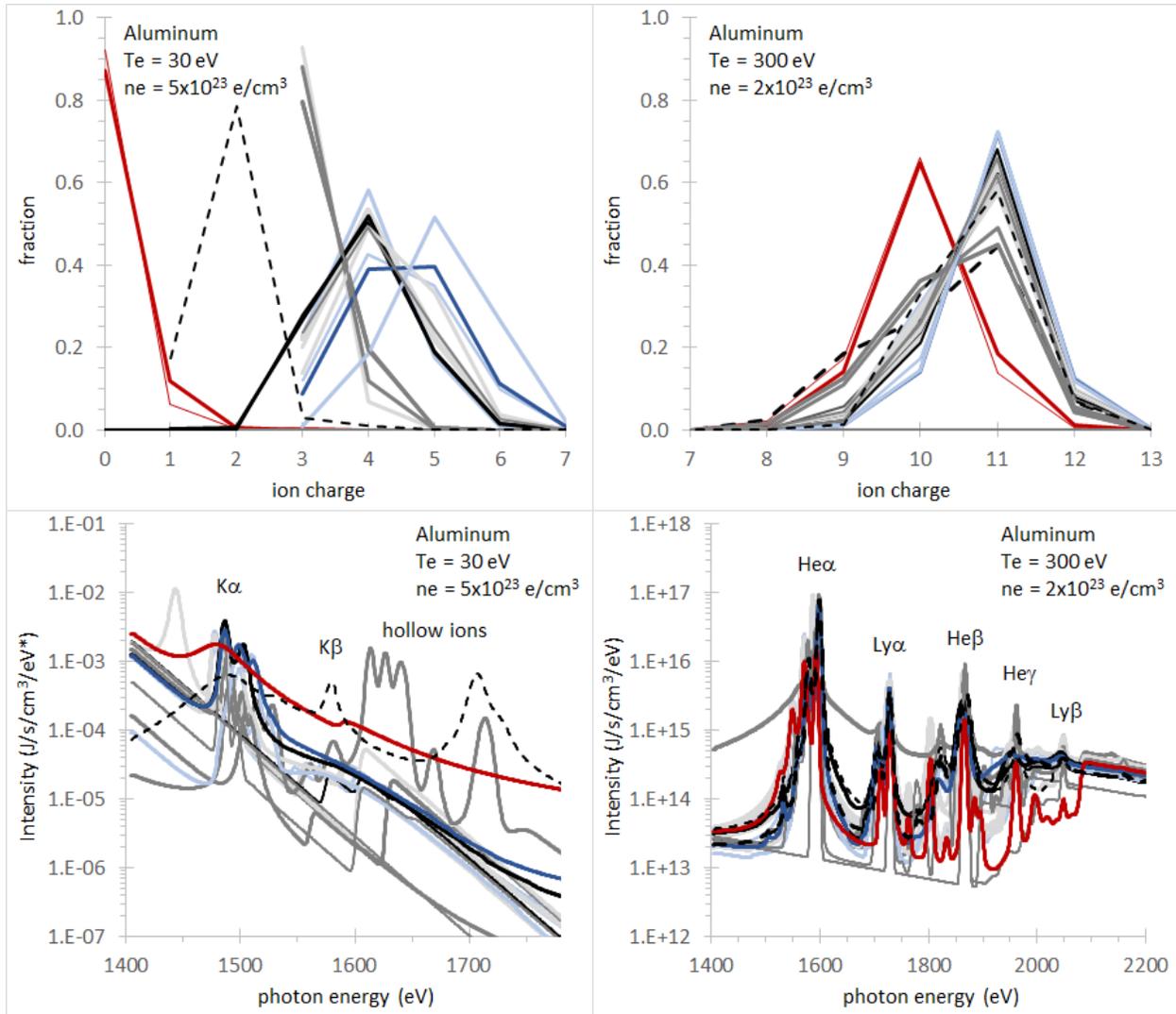

Fig 6. Top: charge state distributions of near-solid-density aluminum for T = 30 eV (left) and 300 eV (right). Bottom: Emission spectra from the two cases, broadened for clarity. Grayscale lines use Stewart-Pyatt or ion-sphere continuum lowering, blue lines use Ecker-Kroll, and red lines exclude continuum lowering effects.

Detailed charge state distributions and emission spectra are given in Fig. 6 for a nearly degenerate case ($T_e$ = 30 eV, $n_e$ =5x10$^{23}$ e/cm$^3$, $E_{Fermi}$ = 16 eV) and a more classical case ($T_e$ = 300 eV, $n_e$ =2x10$^{23}$ e/cm$^3$, $E_{Fermi}$ = 12 eV). The charge state distributions call out the difference between continuum lowering treatments, where models that incorporate fewer or smaller continuum lowering effects have systematically lower charge state distributions. These effects are much more pronounced in the more degenerate case. The emission spectra shown in Fig. 6 include additional 3 eV Gaussian broadening for clarity. In the more classical case (at right), there is fairly good agreement in both absolute and relative intensities among all models, and line energies agree to within a few eV. Here, the most profound differences among the spectra are the location of the modeled edges and the shapes of the far line wings, which reflect best-effort collisional and Stark broadening and are unaffected by the imposed Gaussian broadening. In the near-degenerate case at left, by contrast, there is a much larger spread in absolute and relative intensities (in fact the two models with significant emission features above 1700 eV had to be scaled down by factors of ~10 to fit within the envelope of other codes). Here, most models that use a version of Ecker-Kroll have more Kα emission from higher charge states, in accordance with their higher degree of ionization, and show no Kβ emission, since their M-shell states are completely destroyed. Several of the highly averaged codes with approximate hydrogenic atomic structure predicted cold Kα emission at energies 10 – 50 eV below the reference Ka energy of 1487 eV. Overall, the steady-state aluminum cases indicate that there remains significant uncertainty in the implementation of density effects, and this uncertainty should be considered when using collisional-radiative models to draw out implications of precision spectroscopic diagnostics at extreme densities.

**Time-dependent Al**

The time-dependent aluminum cases were designed to be relevant to the XFEL experiments of Vinko et al, [33] where an X-ray laser beam at various energies was used to irradiate a thin aluminum foil. Aluminum ions in the foil undergo heating from photoionized and Auger electrons and produce fluorescence emission as radiative decay processes fill the XFEL-induced holes in K-shell states. The major puzzle of the measurements was that the standard version of the superconfiguration model SCFLY [34] could not reproduce the observed intensities of fluorescence emission: even with sophisticated spatial, spectral, and temporal profiles for the XFEL beam intensity, FLYCHK with Stewart-Pyatt continuum lowering systematically underpredicted the intensity of fluorescence emission from high charge states. When a modification of Ecker-Kroll continuum lowering was implemented in SCFLY, it produced very good agreement with the measured data at multiple XFEL beam energies. The time-dependent Al case was designed to explore the robustness of intensity variations in the observed fluorescence emission of XFEL-irradiated Al with a wider variety of models and continuum lowering treatments.

As with the neon case above, the temperature evolution was prescribed by SCFLY, starting at 1 eV and ending around 100 eV for both cases. For both cases, the electron densities follow the modeled ionization from the codes at a fixed ion density of $6 \times 10^{22}$ ions/cm$^3$ for solid Al and the XFEL beams were treated as square pulses in time (80 fs) and energy (4.4 eV bandwidth). For the cases with beam energies of 1580 and 1650 eV, respectively, the incident x-ray fluxes were 1.17 and 0.934 x $10^{17}$ W/cm$^2$.

At the lower beam energy, the XFEL photons can directly photoionize Ne-like Al IV/Al$^{3+}$, which is the dominant charge state of Al in ambient conditions with a K-edge at 1560 eV in the cold solid. This photoionization leads to either Auger decay or fluorescence Kα emission from Al V/Al$^{4+}$ at 1487 eV. Depending on the continuum lowering treatment and the details of the atomic structure, the 1560 eV beam may be able to directly photoionize K-shell electrons from higher charge states as well. With the 1650 eV XFEL beam, direct K-shell photoionization of ions up to Al VI/Al$^{5+}$ is possible, leading to emission from Al VII/Al$^{6+}$ at 1510 eV, and emission from higher charge states can be enabled by continuum lowering. As with the 800-eV beam incident on neon, K-shell holes with ionization energies above the beam energy may also be produced by dielectronic recombination, however this is expected to be a much weaker process than direct photoionization and to be less important in this high-density case, where the number of multiply excited states is limited by destruction of high-n orbitals, than it was for the neon gas cases.

As for the neon cases, many fewer individual codes contributed to the time-dependent Al cases than to the steady-state cases. For time-dependent Al, results were submitted from 18 variations of 5 independent codes, only one of which included fine structure.

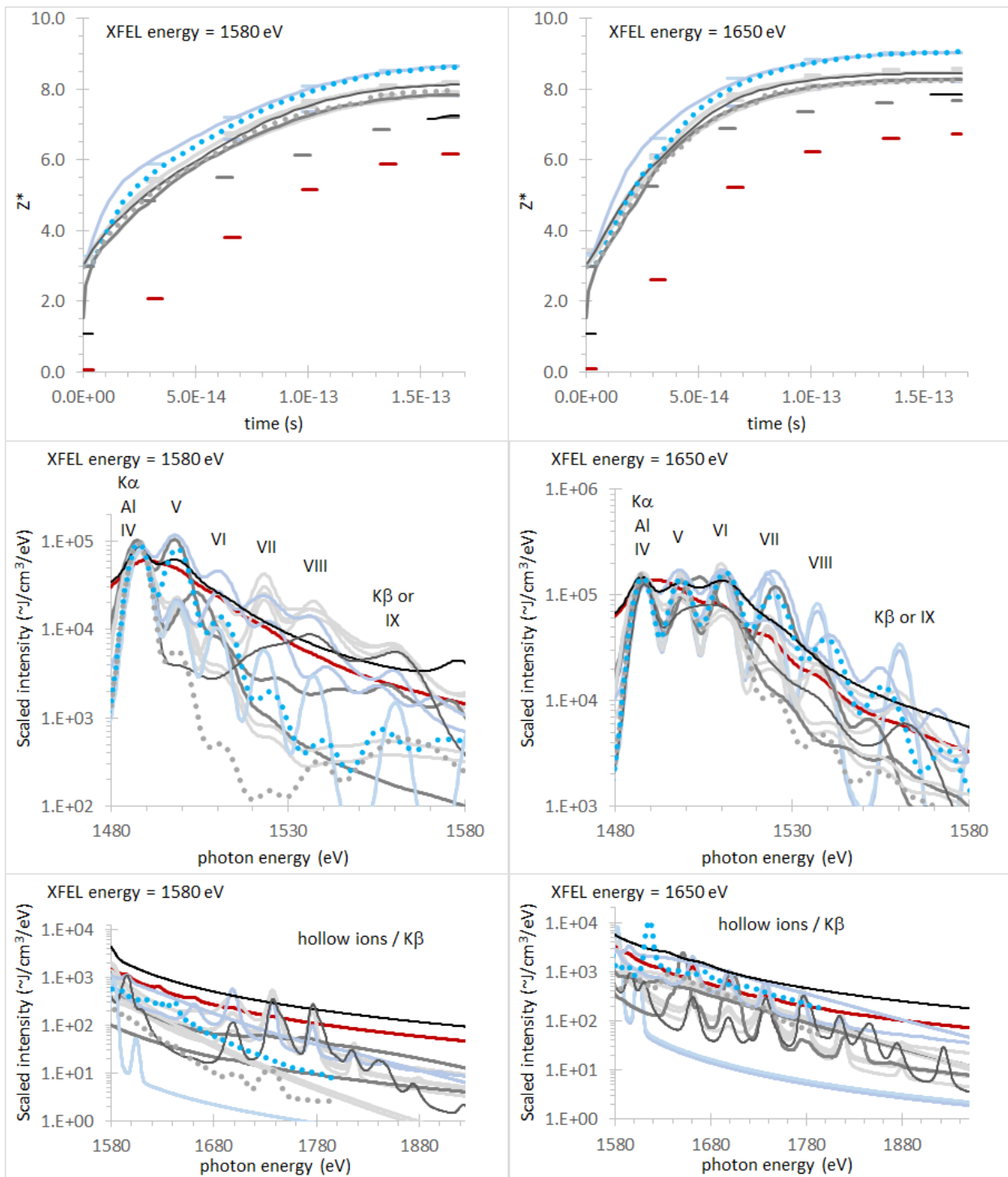

Fig. 7. (color online) Top: Evolution of the average aluminum charge state under irradiation by two x-ray beam energies. Some models provided data only for a subset of the prescribed times and are represented by horizontal dashes. Middle (Bottom): Time-integrated Kα (Kβ) fluorescence spectra. Grayscale lines use Stewart-Pyatt or ion-sphere continuum lowering, blue lines use Ecker-Kroll, and red lines exclude continuum lowering effects. The smooth dotted lines call out SCFLY.

The top panels of Fig. 7 show the predicted time-dependent Z* values for both TD-Al cases. As with the neon cases, some codes reported and/or ran on non-standard time grids; these are indicated by broken lines on the plots of charge state evolution. SCFLY [34], the model used in [32], is called out by dotted lines. As with the steady-state Al cases, rough groupings of Z* with continuum lowering treatment are evident, with all codes predicting more rapid ionization at the higher beam energy. With a given continuum lowering theory, there is agreement among most codes to within about half a charge state among the final Z* values.

The bottom panels of Fig. 7 show the total time-integrated emission spectra from the time-dependent codes. For clarity of presentation, the spectra have been broadened and scaled to the cold K$\alpha$ peaks and the emission peaks are labeled with their parent Al ion. As with the steady-state thermal emission at T = 30 eV, there is a wide variation in model predictions for the relative intensities of emission features from different charge states, in contrast to the relatively good agreement in Z*. While these emission spectra cannot be directly compared to the measured emission from [32,33] because they are not averaged over spatial variations in the XFEL beam intensity, they can help establish the sensitivity of predicted emission features to the structure and density effects implemented in collisional-radiative models.

In the 1580 eV XFEL beam case, where experiments observed higher-than expected K$\alpha$ emission from Al V, clusters of predicted emission with Al V/Al IV K$\alpha$ ratios ranging from 0.1 to 1 are evident. But these clusters do not neatly correspond to the continuum lowering treatment implemented in the models; neither do they obviously correspond to the degree of model detail or completeness as indicated by line shading and thickness. Of the codes that predict the most Al V emission, one is SCFLY with the modified Ecker-Kroll treatment of continuum lowering prescribed in [32] while the others are more detailed models with relatively high rates of dielectronic recombination (the process which gave rise to emission above the XFEL beam energy in neon) enabled by persistent 3$l$ states. Oddly, relatively intense emission from higher charge states is predicted by some of the models with relatively weak emission from Al V.

In the 1650 eV beam case, where experiments observed higher-than expected K$\alpha$ emission from Al VII, the emission intensities from the higher charge states do tend to be grouped roughly according to the continuum lowering treatment. However, one of the more detailed models has relatively intense Al VII emission and several of the more detailed models have relatively indistinct Al V and Al VI features. Finally, there are also wide variations in predicted K$\beta$ and hollow-ion emission features for both beam energies.

**Steady-state Si**

The steady-state Si cases were designed to be relevant to the benchmark photoionization emission and absorption measurements described in [28]. The major puzzle of this experiment was that multiple models predicted a higher ionization for the inferred plasma conditions than was implied by the absorption spectrum. The temperature of the plasma was measured using the ratios of absorption features from the ground (1s$^2$2s) and excited (1s$^2$ 2p) states of Li-like Si$^{11+}$. If those states are in collisional equilibrium (with populations proportional to $g_i e^{-E_i/T_e}$) and the oscillator strengths of K-shell photoexcitation from those states to 1s 2$l$ n$l$ are accurately known, then the ratio of their absorption features should give a direct measure of the electron temperature $T_e$. For the models used in the paper, the inferred $T_e$ was 33 eV and both the spectrum and intensity of the driving radiation field were measured. The expansion of the sample was measured, giving an estimate for the electron density of 8.5x10$^{19}$ e/cm$^3$. The measured absorption spectra indicated charge state distributions that peak around Be-like Si$^{10+}$, while the measured emission was dominated by Li-like Si$^{11+}$. However, both the ATOMIC [12] and XSTAR [40] models used to model the emission in [28] predicted emission and absorption dominated by He-like Si$^{12+}$. To fit the measured spectra, both models had to decrease the temperature and/or increase the density to enhance collisional recombination. The Si case in the NLTE workshop was designed to test the robustness of this discrepancy with a wider variety of models.

While both external radiation fields and self-photopumping due to optically thick lines have been explored in previous workshops, this was the first time that both non-Planckian external fields and combined effects of optical depth and external radiation were considered. This case had among the fewest submissions of all steady-state cases, with 14 variations of 10 codes providing results. Most of the codes included fine-structure detail, which was required to match the extremely high-resolution experimental data. The submitted models spanned an enormous range of statistical completeness, with total statistical weights ranging from $10^4$ to $10^{10}$.

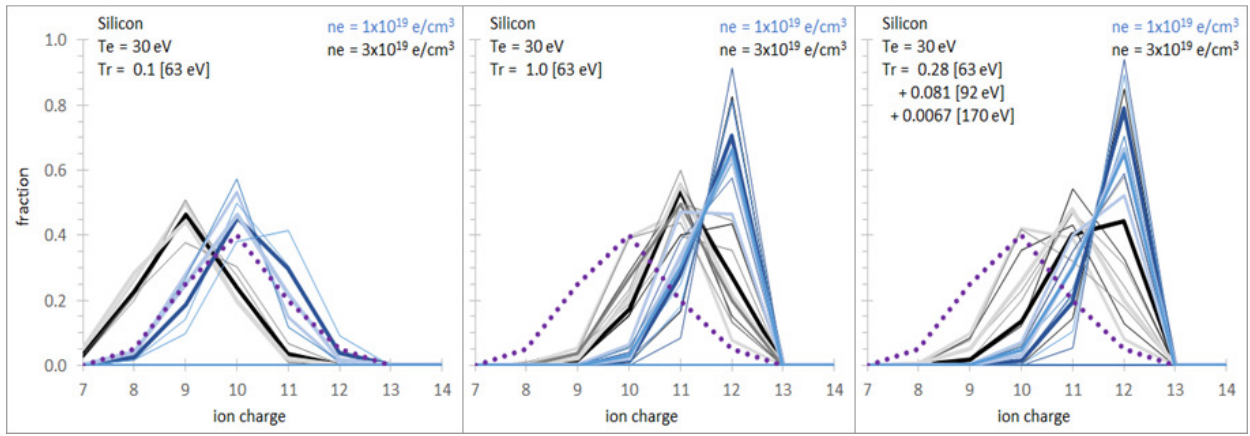

Fig. 8: (color online) Charge state distributions for Si plasmas with $L$ = 0.1 cm, $T_e$ = 30 eV, two electron densities (blue and black for 1 and 3 x $10^{19}$ e/cm$^3$, respectively) and three different radiation fields. Dotted lines represent a good fit to experimental emission and absorption data from [28].

Figure 8 shows modeled charge state distributions for six sets of plasma conditions along with a dotted line representing a good fit to the experimental data from [28]. All results here are from the thinnest plasma (0.1 cm in length) at the lowest requested temperature (30 eV). Results are given for two densities (blue and black for 1 and 3 x $10^{19}$ e/cm$^3$, respectively) and three radiation fields: a diluted Planckian at radiation temperature Tr = 63 eV, a Planckian at Tr = 63 eV, and a multi-Planckian field composed of different fractions of three color temperatures (0.28 of 48 eV, 0.081 of 92 eV, and 0.0067 of 170 eV). The full 63 eV Planckian and the multi-Planckian fields have similar total energy fluxes, with the multi-Planckian having significantly more photons at higher energy. For all radiation fields, the higher electron density leads to lower charge states due to increasing rates of collisional recombination (here dominated by dielectronic recombination) relative to the field-driven photoionization rates. And while most models predict the same most probable charge state for each of the four cases summarized here, there is significant scatter in the details of the charge state distributions, particularly for the multi-Planckian radiation field. The models predict ionization increases of about 2 charge states moving from the diluted Planckian to either the undiluted or multi-Planckian external field. At the case conditions closest to the experimentally inferred parameters (33 eV, 8.5x$10^{18}$ e/cm$^3$ with either radiation field), none of the models have charge state distributions that would give rise to absorption spectra dominated by Be-like Si$^{10+}$, as was observed in the experiment. Even increasing the density to ~3.5x the inferred value, none of the codes predict charge state distributions matching the one that fit the experimental absorption spectrum. However, most of the codes do roughly match the measured CSD at $n_e$ =$10^{19}$ e/cm$^3$ with the diluted Planckian.

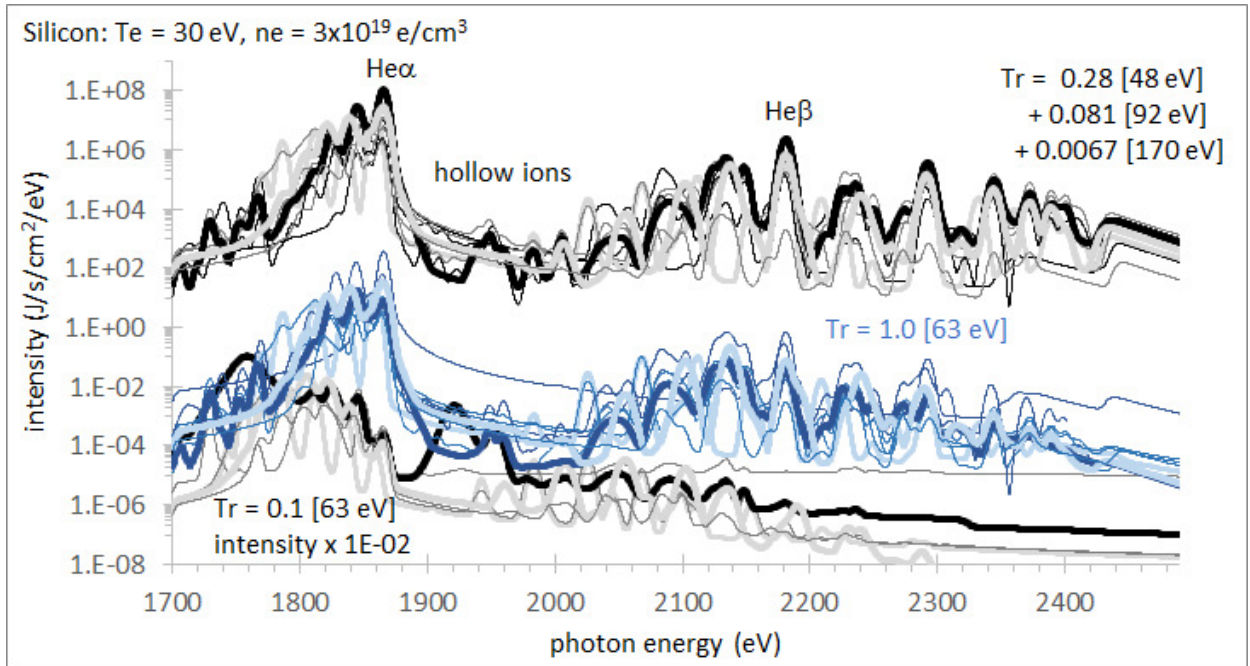

Fig 9. (color online) Emission spectra from Si plasmas with L=0.1 cm, $T_e$ =30 eV, $n_e$ = $3 \times 10^{19}$ e/cm$^3$, and three different radiation fields.

Figure 9 shows a collection of emission spectra from the 30 eV, $3 \times 10^{19}$ e/cm$^3$ case for the three radiation fields. The spectra have been broadened and the intensities of the diluted Planckian emission have been reduced by a factor of 0.01 for clarity, but are otherwise unmodified and so accurately represent the differences in absolute modeled intensities. While the undiluted and multi-Planckian fields had similar impact on the predicted Z* values, the multi-Planckian has a much larger effect on the emission intensity compared to the undiluted Planckian. This is because the charge state distribution is driven by photoionization of L-shell electrons by few-hundred-eV photons, where both radiation fields have similar intensities. By contrast, the K-shell emission is driven by photoionization of K-shell electrons by photons with energy >1839 eV, and at those energies the multi-Planckian radiation field has >$10^6$ more photons. Overall, there is fair agreement among models in the absolute emission intensities and spectral features, although one detailed and complete model predicts anomalously intense hollow-ion emission for the diluted Planckian. More importantly, the relative intensities of lines from different charge states roughly follows the charge state distributions for all models, suggesting that the spectra are reliable indicators of Z*.

Comparisons of detailed and unbroadened emission spectra from the most detailed models at the lowest temperature, highest density, and two radiation fields are given in the top panel of Fig. 10 for the thinnest plasma, scaled to the Li-like satellite intensities with vertical offsets for clarity. As with neon and Al, there are few-eV variations among models in line energies and a fairly wide range of emission envelopes and intensities for the satellite features from L-shell charge states. For the Planckian radiation field, most models predict dominant emission from Li-like Si and fairly weak emission from the He-like resonance and intercombination line. With the multi-Planckian field, most models predict significant enhancements of these lines, even if their charge state distributions are similar for the two fields. This is due to direct photopumping of the He-like resonance line and radiative cascades from other photopumped excited states into the metastable state that gives rise to the intercombination line. In the bottom panels of Fig. 10, the emission from a thick plasma at the same temperature, density and fields is presented. Here, opacity effects change the envelopes and relative intensities of emission from different charge states for both radiation fields. And in the multi-Planckian, optically thick case, the models that include self-photopumping of excited states as well as absorption along the line of sight (about half the models shown) show a significant enhancement in the intercombination line intensities, again due to radiative cascades from enhanced populations in the excited states. We note that while the models here treat the plasma as an infinite slab at the given depth (with a mean photopumping length twice the slab thickness), the experimental plasma geometry is not far from square (with a mean photopumping distance smaller than the given thickness) at the time of the absorption and peak emission measurements, so the modeled geometry probably overestimate the effects of self-photopumping that might affect the experiment.

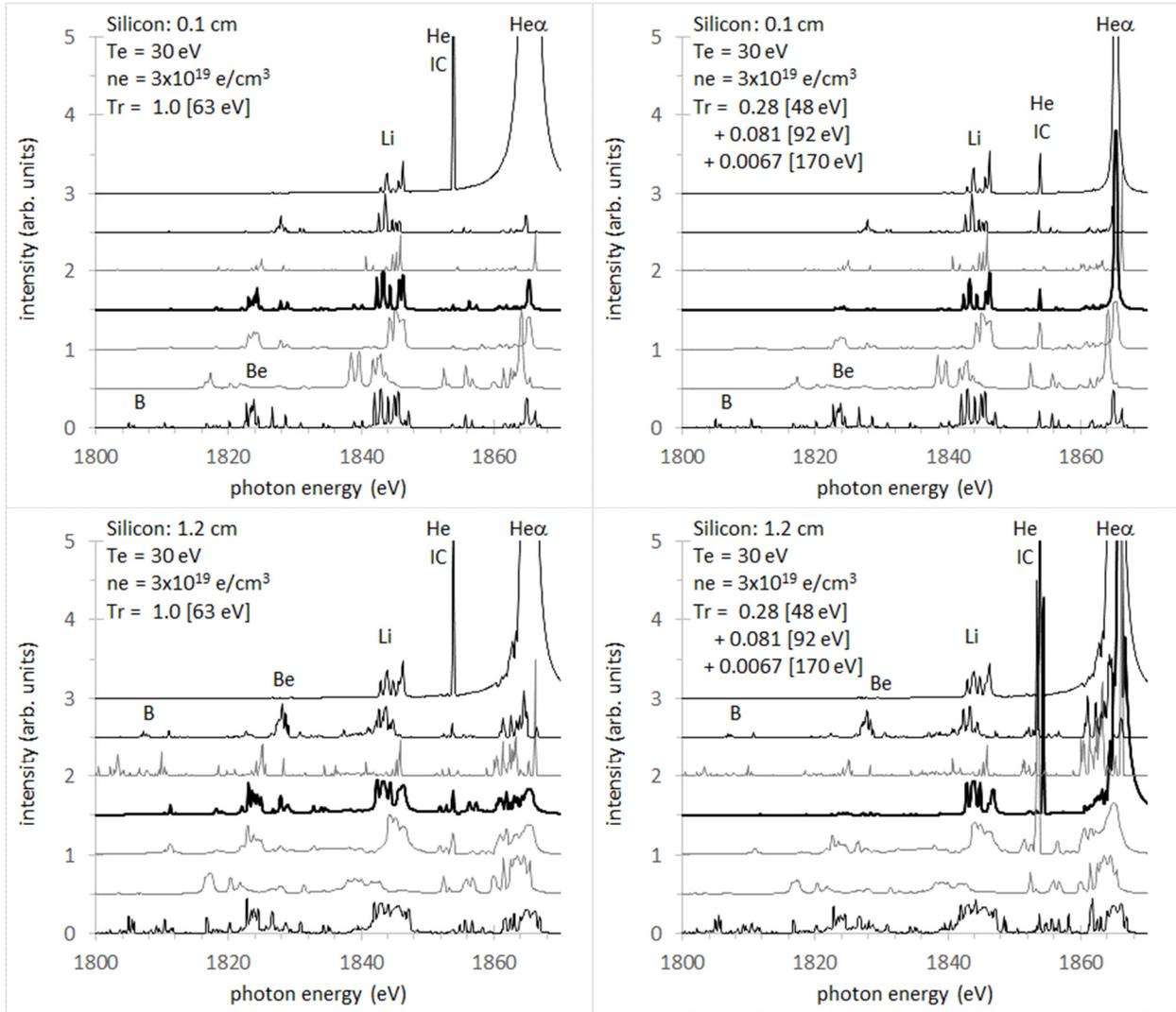

Fig 10. Detailed Si line emission at various conditions.

As noted above, the experiments used the ratio of absorption features from ground and excited states of Li-like Si to determine a plasma electron temperature of 33 +/- 7 eV. While absorption spectra have not historically been explored at the non-LTE workshops, we invited participants to submit them for this case in a call for resubmission issued after the workshop was held. In a further break from tradition, we also invited calculations at additional temperatures (60 and 100 eV) and with an additional radiation field (the diluted Planckian). Expanded resubmissions were provided for only 7 variations of 5 codes; only four of which had fine-structure detail, and only two whose submitted data sets included both absorption and the additional temperatures. Figure 11 shows the results for Li-like absorption from those codes for the higher density and Planckian radiation field (similar results were observed for the lower density and the multi-Planckian). While there are differences in line shapes and energies among all three codes, both of the codes that submitted absorption data for the higher temperatures show an increase in the absorption depth of the excited-state (E.S.) relative to the ground-state (G.S.) lines as expected for excited-state populations that increase with temperature. But at the lowest temperature, the ratio of the two strongest G.S. and E.S. lines differs significantly among the three codes, and while the experimentally observed ratio would fit the black-line model at 30 eV, the other two models might be expected to reproduce the observed ratios only at lower temperatures where excited states are less populated. The plot on the right of Fig. 11 shows the populations of the $1s^2\,2p$ states in all the models at a temperature of 30 eV and with the Planckian distribution at both $1 \times 10^{19}$ e/cm$^3$ (open circles) and $3 \times 10^{19}$ e/cm$^3$ (solid circles), color-coded to the spectra on the left. Most of the models predict $1s^2\,2p$ populations above those expected in LTE, particularly at the lower density (closer to the inferred plasma density), indicating that the satellite-line-based temperature diagnostic may be more sensitive than expected to the driving radiation field. There also appears to be more than expected variation in satellite-line oscillator strengths, given the disagreement in satellite absorption intensities for models with similar populations. Although this analysis was limited to only a few models, it suggests that populations, oscillator strengths, internal collision strengths that drive populations to LTE, and line shapes may all impact the interpretation of the E.S. to G.S. line ratio as a temperature diagnostic. While additional effects such as time-dependence in level populations and experimental conditions may be needed to ultimately resolve the discrepancy between the experimentally inferred CSD and the general model consensus of a higher CSD at the inferred conditions, this sensitivity of the thermometer may be an important consideration.

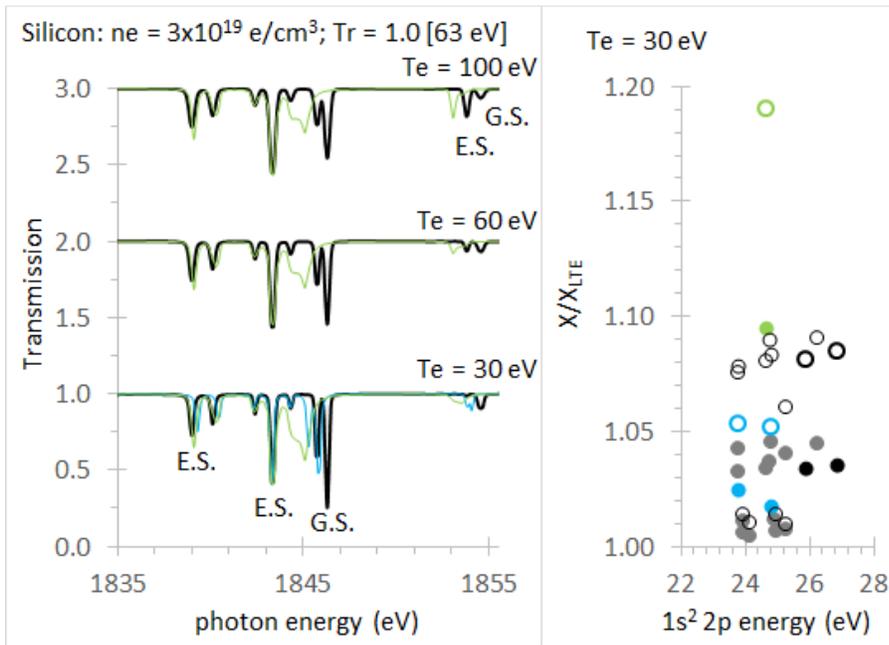

Fig. 11. (color online) Left: Temperature dependence of the Li-like satellite absorption reflecting increasing occupation of the $1s^2\,2p$ excited states (E.S.) in Li-like Si. Right: Ratios of modeled occupations for the $1s^2\,2p$ states in Si with the Planckian radiation field, $T_e = 30$ eV, and $n_e = 10^{19}$ e/cm$^3$ (open circles) and $3 \times 10^{19}$ e/cm$^3$ (filled circles).

**Steady-state Cl**

The chlorine cases were designed to be relevant to recent measurements of Cl Heβ and its Li-like satellites using the high-resolution OHREX spectrometer generated from chlorinated plastic targets irradiated by both long- and short-pulse beams on ORION [27]. The measured He- and Li-like line widths and relative emission line intensities were used to infer a temperature of 500 eV and a density of $5\times10^{22}$ e/cm$^3$ in the experiment. Since collisions among n =3 states drive both the Heβ line widths and the population distributions that control the relative intensities of the Li-like satellite lines, this case, like the neon and aluminum cases, was useful in exploring density effects and their application to plasma diagnostics. In addition to the standard grid of conditions, participants were also invited to provide their code's best-fit to the experimental data from [27].

Results for the Cl cases were submitted from 26 variations of 16 independent codes, about half of which were sufficiently detailed to fit the high-resolution experimental spectrum. Two of the detailed codes used sophisticated line-shape models for the Heβ line that included effects such as the Stark-effect-driven appearance of the dipole-forbidden $1s^2$ ($^1$S) – 1s3s ($^1$S) transition, which is spectroscopically indistinguishable from the intercombination line. Other detailed codes used best-effort broadening, most often informed by parameterized Stark effects and collisional widths consistent with the collisional rates internal to the CR models. Statistical completeness varied widely, with total statistical weights ranging from $10^3$ to $10^9$.

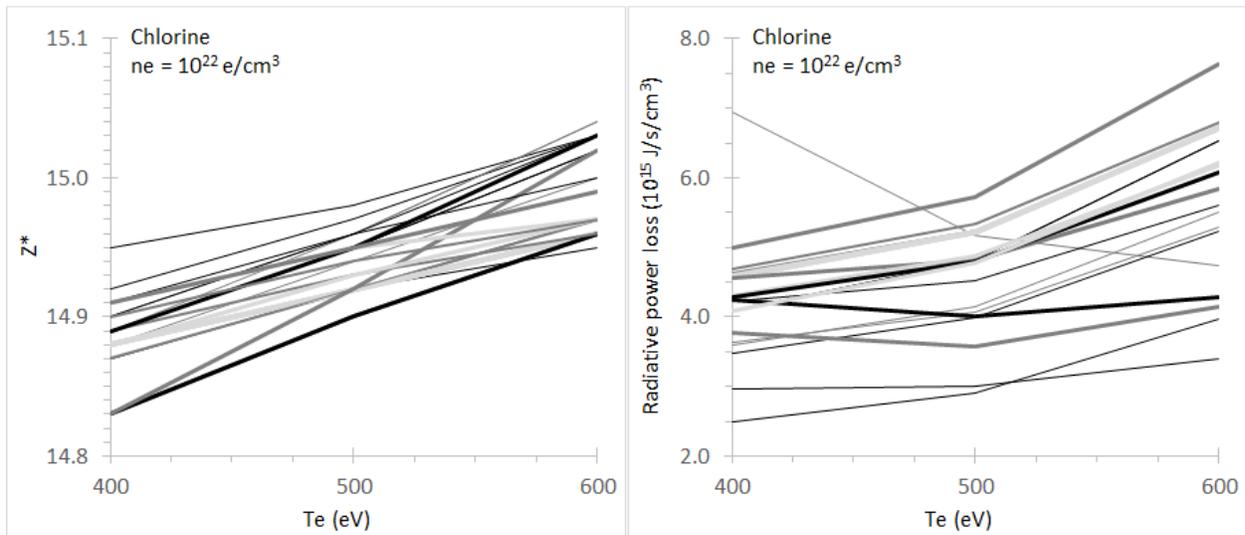

Fig. 12: Left: Average ionization Z* of Cl. Right: Radiative power losses from Cl at the same conditions.

Despite the significant variety of models, remarkable agreement was seen in the calculated values of both the average ion charge and radiative power losses. This agreement is illustrated in Fig. 12, where the majority of codes agree to within ~0.1 charge state in Z* and ~30% in RPL. This agreement carried over to the other densities, even to $10^{23}$ e/cm$^3$, where continuum lowering truncates the bound shells to n <~ 7 (for Stewart-Pyatt) and n <~ 5 (for Ecker-Kroll). The good agreement even in a regime where continuum lowering treatments matter is due to the near-closed-shell configuration (He-like Cl has Z* = 15), where the ~500 eV temperature is well below excitation and ionization energies of 2-3 keV, ensuring that most of the He-like population lies in the ground state and is unaffected by the relatively modest differences in continuum lowering predictions from various theories for either ionization energies or accessible statistical weights in the multiply excited He-like states.

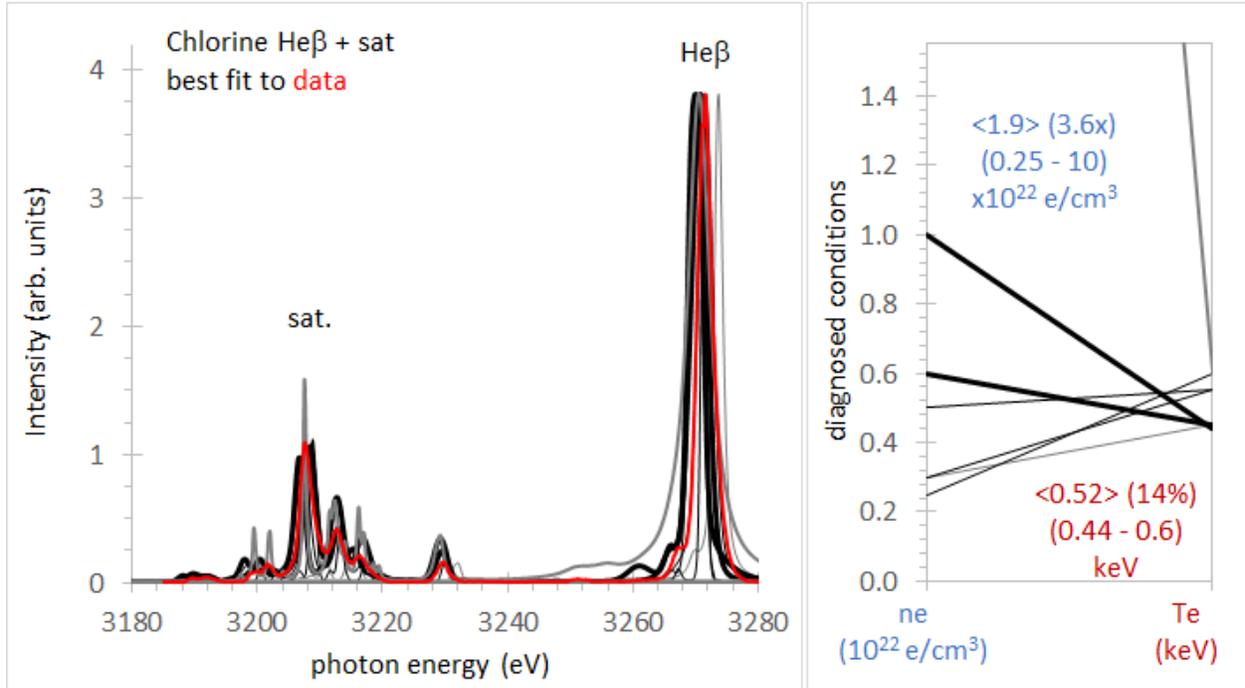

Fig. 13. (color online) Left: Best-fit modeled spectra compared to data from [27], which is given in red. Right: Diagnosed densities and temperatures from best-fit spectra.

The good agreement in Z* and RPL is reflected in the calculated spectra, especially among the more detailed models. The best-fit spectra to the experimental data of [27] from these detailed models are shown in Fig. 13 along with the measured spectrum (in red) and the variations in diagnosed densities and temperatures. While there are few-eV differences in line energies and significant variations in the width of the Heβ line, the overall envelope of the satellite features is well-matched by all models. The relative satellite intensities are driven by density-dependent internal collisions among 1s2*l*3*l′* autoionizing states, and most models that used these satellite features to infer the electron density diagnosed $n_e$ to be between 0.2 and 1 x$10^{22}$ e/cm$^3$ (60% standard deviation). One model without detailed satellites inferred a significantly higher density of $10^{23}$ e/cm$^3$. The Heβ width is affected by Stark broadening and opacity effects as well as collisions, and is well-fit by only a few of the models. The temperature is inferred primarily from the relative intensity of the Heβ resonance and Li-like satellite features; this ratio is also affected by the opacity. Even so, the variation in the diagnosed temperature is rather small, with only a 14% standard deviation among models. These results are consistent with those of the best-fit Kr L-shell case described in [41], where factors of 2-3 in density and ~20% in temperature are asserted as typical for spectroscopic fitting with detailed CR models.

**Discussion**

Most of the cases in the 10$^{th}$ NLTE workshop touched directly on recent experiments from diverse HED facilities including X-ray free electron lasers, high-power optical lasers, and pulsed-power-driven plasma and radiation sources. These investigations provided insight into the reliability (as measured by agreement among diverse collisional-radiative codes) of widely-used spectroscopic diagnostics. In general, we found fair agreement among models in the lower-density plasma cases, even after introducing complexities such as external radiation fields (silicon) or time-dependence under x-ray beam irradiation (neon). At higher electron densities, fair agreement among diverse models was also evident in thermal and non-degenerate aluminum and chlorine cases, albeit with significant uncertainty remaining in details such as line shapes. The largest disagreement among models in this workshop was the high-density, near-degenerate aluminum case, where wide variations in the predicted emission spectra from both steady-state and time-dependent conditions appear to reveal the fundamental incoherence of pushing collisional-radiative models based on isolated-atom atomic structure into regimes where density effects such as continuum lowering and degenerate electron energy distributions introduce order-of-magnitude changes to model structure. In such regimes, CR models appear to be extremely sensitive not only to the theories used to implement ad-hoc density corrections (e.g. Stewart-Pyatt or Ecker-Kroll) but also to idiosyncrasies of their underlying structure and the precise implementations of the density effects. Future workshops will continue to explore these challenging regimes, and we especially welcome additional data collected from modern high-energy-density facilities that can help constrain and stimulate development of CR models better suited to the extreme environments now accessible in XFEL and ICF experiments.

**Acknowledgements**

We thank Farhat Beg and Meghan Murphy of UCSD for help with local organizing and are grateful to all of the contributors and speakers who participated in this workshop. Sandia National Laboratories is a multimission laboratory managed and operated by National Technology and Engineering Solutions of Sandia, LLC., a wholly owned subsidiary of Honeywell International, Inc., for the U.S. Department of Energy's National Nuclear Security Administration under contract DE-NA-0003525. The work of SBH was supported by the U.S. Department of Energy, Office of Science Early Career Research Program, Office of Fusion Energy Sciences under FWP-14-017426. The work of CJF was performed under the auspices of the United States Department of Energy under contract DE-AC52-06NA25396. The work of HAS was performed under the auspices of the United States Department of Energy under contract DE-AC52-07NA27344. The work of ES was supported in part by ONRG (USA).